\documentclass[oribibl]{llncs}

\usepackage{amsmath,amsfonts,amssymb}
\usepackage{thm-restate}
\usepackage{hyperref}
\usepackage{tikz}
\usetikzlibrary{arrows,automata}
\usetikzlibrary{calc}
\usepackage[noend]{algorithm2e}
\usepackage{stmaryrd}
\usepackage{macros}
\usepackage{graphicx}
\usepackage{microtype}

\newif\iffinal
\finalfalse

\newcommand{\myorcid}[1]{\unskip\texorpdfstring{%
\href{https://orcid.org/#1}{\includegraphics[width=10px]{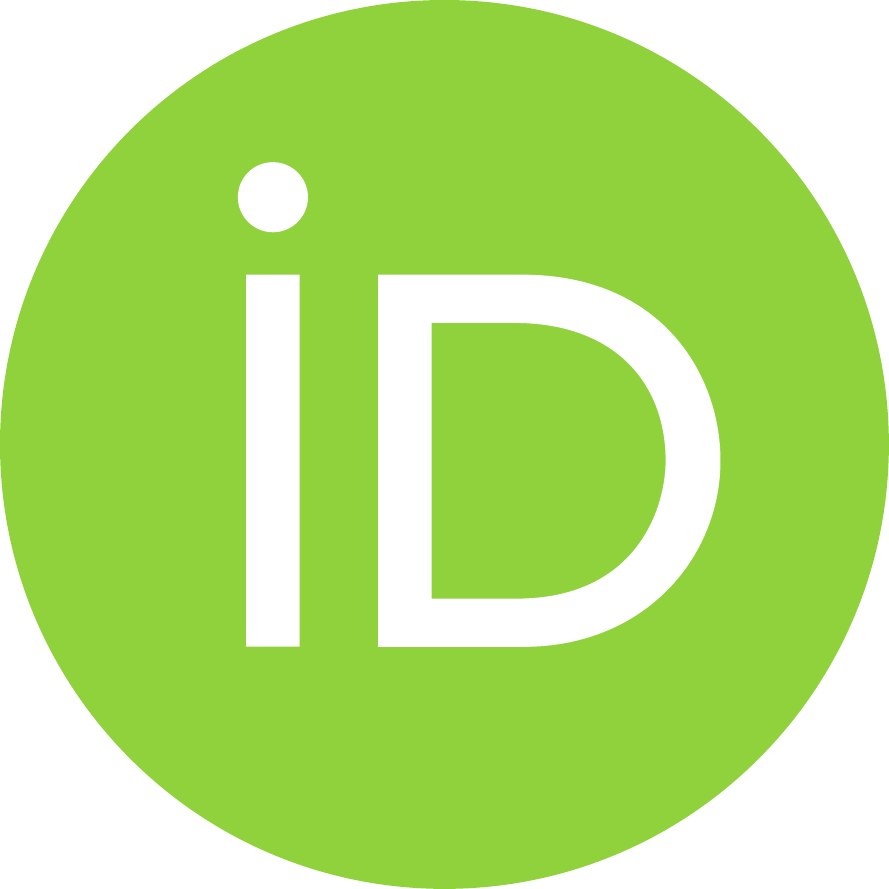}}}{}}

\begin{document}

\title{Incremental methods\\
  for checking real-time consistency%
  \thanks{This work was partially funded by ANR project Ticktac
    (ANR-18-CE40-0015), and by a MERCE\slash Inria collaboration.}}

\iffinal 
\author{Thierry~J\'eron\inst{1} \myorcid{0000-0002-9922-6186}
  \and
  Nicolas~Markey\inst{1}      \myorcid{0000-0003-1977-7525}
  \and
  David~Mentr\'e\inst{2}       \myorcid{0000-0003-4315-0335}
  \\
  Reiya~Noguchi\inst{2}        
  \and
  Ocan~Sankur\inst{1}          \myorcid{0000-0001-8146-4429}
  }
\else 
\author{Thierry~J\'eron\inst{1} 
  \and
  Nicolas~Markey\inst{1}      
  \and
  David~Mentr\'e\inst{2}       
  \\
  Reiya~Noguchi\inst{2}        
  \and
  Ocan~Sankur\inst{1}         
  }
  \fi  
\institute{Univ Rennes, INRIA, CNRS, Rennes (France)\\
   \email{firstname.lastname@inria.fr}
     \and Mitsubishi Electrics R\&D Centre Europe, Rennes (France)\\
   \email{initial-of-firstname.lastname@fr.merce.mee.com}}
\maketitle
\begin{abstract}
  Requirements engineering is a key phase in the development process.
  Ensuring that requirements are consistent is essential so that they
  do not conflict and admit implementations.  We~consider the formal
  verification of \emph{rt-consistency}, which imposes that the inevitability
  of definitive errors of a requirement should be anticipated, and
  that of \emph{partial consistency}, which was recently
  introduced as a more effective
  check.  We~generalize and formalize both notions for discrete-time timed
  automata, develop three incremental algorithms, and present
  experimental results.
\end{abstract}

\section{Introduction}
In the process of developing computer systems, requirement engineering consists in defining, documenting and maintaining the requirements.
Requirements can be of different nature, 
but since we are interested in timed systems, i.e. systems where time constraints are of importance, we will focus here on timed functional ones.
Requirements are the primary phase of the development process,
and are used to partly drive the testing campaign in order to check that they are indeed satisfied by the implementation.
In a formal approach, it is thus important to design formal requirements that are consistent,
\emph{i.e.} that are not contradictory and admit implementations that conform to them.

In this paper, we study two prominent consistency notions studied in
the literature for real-time system requirements, called
\emph{rt-consistency}~\cite{Post2011} and \emph{partial
  consistency}~\cite{Becker2019}.
Partial consistency concentrates the notion of consistency on Simplified Universal Patterns
(SUP)~\cite{bienmuller2016modeling} which are simple real-time
temporal patterns used to define real-time requirements, essentially
comprising an assumption (named \emph{trigger}), a~guarantee (named
\emph{action}), together with timed constraints on delays of these and
between them.
The~advantage of SUPs is that they define a specification language that is expressive enough yet easy to
understand, even by non experts.
The counterpart is that the notion of partial consistency is specific to them and tricky.

Rt-consistency requires that all finite executions that do not violate the requirements,
have infinite extensions that satisfy all of requirements.
Put differently, this means that if an implementation produces a finite
execution whose all continuations necessarily lead to the violation of
some requirement, then there must be a requirement that is already violated by the finite execution.
In simple words, inevitability of errors should be anticipated by the set of requirements.
Thus, rt-consistency ensures that the set of requirements is well designed and sane.
This is interesting in that it may reveal conflicts between requirements and catch subtle problems, but it is rather expensive to check.
Several directions can be investigated to mitigate this complexity: restrict to sub-classes of requirements, in particular SUPs, restrict to subsets of requirements, examine alternative and cheaper notions of consistency.
However these lead in general to false positives and false negatives, and avoiding them requires additional conditions or checks. 

Partial consistency is one of these alternative notions of consistency 
that only considers pairs of
SUP requirements. It checks that
if there are possibly different executions that trigger both requirements
and satisfy one of them, then there should be a common execution in which both 
requirements are triggered and satisfied.
This check is perhaps better understood as a necessary condition for the rt-consistency of \emph{subsets} of requirements
(but this does not imply the rt-consistency of the whole set).
We formalize this link in this paper.
The general motivation is to gain in efficiency, both by restricting to pairs of requirements,
but also by focusing on particular situations where inconsistencies may arise.
Nevertheless partial consistency can still be costly to check.

\paragraph{Contributions.}
We address the efficiency issue mentioned above by considering an incremental approach to checking consistency and finding inconsistencies in real-time requirements.
In fact, rt-consistency and (bounded) partial consistency are rather expensive to check already on small examples, and because of the state-space explosion problem (which is a classical problem when composing several systems or properties), there is no hope that the approaches would scale to large sets of requirements.
Our algorithms improve the scalability of this approach by allowing one to check larger sets of requirements. We also define a new notion of incremental consistency, and allow to get different degrees of confidence about consistency (up~to full rt-consistency).

We show that checking rt-consistency can be
reduced to \CTL
model checking for discrete-time systems, providing an
alternative approach to duration calculus and timed automata model
checking of~\cite{Post2011}.
Then, we~develop incremental algorithms
for checking rt-consistency and a variant of partial consistency generalized for automata.
Inconsistencies are searched by starting with small batches of requirements.
Whenever we find a counterexample to consistency, we either
confirm~it (by~checking that it fulfills the other requirements) or start the
analysis again with more precision by adding a new requirement in the
batch. This helps us to scale our analysis to larger sets of
requirements.
This idea is applied separately for both consistency notions.
Moreover, we formalize the relation between the two notions, showing
how to obtain counterexamples to rt-inconsistency from counterexamples to partial consistency.
\iffinal
\else
Due to space constraints, all proofs are given in the appendix. 
\fi

\paragraph{Related works.}
Consistency notions appear naturally in the contract-based design of
systems~\cite{BenvenisteCNPRR18}. In~this setting, consistency is
defined as the existence of an implementation of a contract, which
relates environment and system behaviors via assumptions and
guarantees. The related notion of \emph{existential consistency} is
studied in~\cite{ESH-fmics2014}, where consistency consists in the
existence of an execution satisfying the requirements.

Simplified Universal Patterns were introduced in
\cite{bienmuller2016modeling} to simplify the writing of requirements
by non-experts. The patterns are in the form of an assumption and
guarantee.  In~this paper, the~notion of consistency ensures the
existence of an execution which realizes one requirement (both the
assumption and the guarantee) without violating any other one.
In~\cite{bienmuller2016modeling}, the authors also use coverage
notions to measure sets of consistent executions to give a
quantitative measure of consistency.  The notion considered there is
thus related to \emph{non-vacuity} (see e.g.~\cite{PHP-irec2011}).

More reactive notions were studied as in~\cite{AHLNT-sttt2017} where
consistency requires that the system should react to uncontrollable
inputs along the execution so as to satisfy all requirements. The
notion is thus formalized as a game between the system and the
environment, and an SMT-based algorithm is given to check consistency
within a given bound.  This notion thus relies on alternation of
quantifiers at each step.  Rt-consistency and partial consistency, which we~consider in this paper,
lie between the two extreme approaches (that~is simply
existential \emph{versus} game semantics).  In~fact, a~single
quantifier alternation is needed to define rt-consistency (see
Section~\ref{sec:rt-consistency}).  The~rt-consistency checking
algorithm of~\cite{Post2011} considers systems in a continuous-time
setting, and uses duration calculus and timed automata model
checking. We~consider discrete-time systems (with unit delays rather
than arbitrary real-valued delays).

\section{Definitions}
\subsection{Computation Tree logic}

We use \CTL to characterize certain kinds of
inconsistencies. \CTL formulas are defined as
\(
\CTL \ni \phi \coloncolonequals p \mid \neg \phi \mid \phi\vee\phi \mid
\A \X \phi \mid \E\G\phi \mid \E\phi\U\phi,
\)
where $p$ ranges over~$AP$.
\CTL formulas are evaluated at the root of computation trees.
We thus consider computation trees labeled by valuations of atomic propositions: a~tree~$t$
is a set of finite non-empty traces, i.e. words over $2^{\AP}$, closed under prefix, hence containing
exactly one trace of size~$1$ (called its root, and denoted with~$r(t)$).
We~denote~$\prefix$ the prefix ordering on traces. 
Given a node in the tree represented by a trace
$\sigma\in t$,  we~write $t_{\sigma}$ for the subtree of~$t$
rooted at~$\sigma$ (i.e., the set of all traces~$\sigma'$ such
that $\sigma\cdot\sigma'\in t$).
We~write~$\sigma[i]$ for the prefix of length~$i$ of~$\sigma$.
That a tree~$t$ satisfies a formula~$\phi\in\CTL$ is defined
as follows:
\begin{xalignat*}3
 t&\models p &\iff& & &p \in r(t)(p)\\\noalign{\pagebreak[3]}
 t&\models \neg\phi &\iff& & &t\not\models\phi \\\noalign{\pagebreak[3]}
 t&\models \phi\vee\phi' &\iff& & &t\models\phi\text{ or }t\models\phi' \\
t&\models \A \X \phi  &\iff&&& \forall \sigma\in t.\
(t_{\sigma[1]}\models \phi)\\
t&\models \E\phi\U\phi' &\iff&&& \exists \sigma\in t.\
(t_{\sigma}\models \phi' \text{ and }
  \forall\sigma'.\  (r(t)\prefix\sigma'\prefix\sigma) \Rightarrow
  t_{\sigma'}\models\phi)\\
t&\models \E\G\phi&\iff&&& \exists \sigma\in t.\
  (\forall i.\ t_{\sigma[i]}\models \phi)
\end{xalignat*}
Using $\A\X$, we can define $\E\X$ by $\E\X\phi \equiv \neg \A\X\neg
\phi$.  Similarly, $\A\F\phi \equiv \neg \E\G\neg\phi$ means that
$\phi$ holds along any infinite branch of the tree, and finally $\A\phi\U\phi'
\equiv \A\F\phi' \et \neg\E(\neg\phi')\U(\neg\phi\et\neg\phi')$
means that along all infinite branch, $\phi'$ eventually holds and
$\phi$ holds at all intermediary nodes.

\subsection{Timed automata}
\label{sec:ta}
We consider requirements expressible by a class of \emph{timed automata} (TA)~\cite{AD90}.
These extend finite-state automata with variables, called
\emph{clocks}, that can be used to measure (and impose constraints~on)
delays between various events along executions.
More precisely, given a set $\Cl=\{c_i \mid 1\leq i\leq k\}$ of clocks,
the set of \emph{clock constraints} is defined by the grammar:
\(
g \coloncolonequals c \sim n \mid g\et g
\),
where $c \in \Cl$, $n \in \bbN$, and $\mathord\sim \in \{\mathord<, \mathord\leq, \mathord=,\mathord\geq,\mathord>\}$.
Let $\ClC(\Cl)$ denote the set of all clock constraints.

We~consider integer-valued clocks 
whose semantics of constraints is defined in the expected way: given a clock valuation
$v\colon \Cl \to \bbN$, a constraint~$g\in\ClC(\Cl)$ is true at~$v$, denoted $v\models g$, 
if the formula obtained by replacing each occurrence of~$c$ by~$v(c)$ holds.
For a valuation~$v\colon \Cl \to \bbN$, an
integer~$d\in\bbN$, and a subset~$R\subseteq\Cl$, we define
$v+d$ as the valuation~$(v+d)(c) = v(c) + d$ for all~$c \in \Cl$,
and~$v[R\leftarrow 0]$ as $v[R\leftarrow 0](c) = 0$ if~$c \in R$,
and~$v[R\leftarrow 0](c) = v(c)$ otherwise.
Let $\bfzero$ be the valuation mapping all variables to~$0$.

We consider timed automata as
monitors of  the evolution of the system through the observation
of values of Boolean variables.
We~thus consider a set $\AP=\{b_i \mid 1\leq i \leq n\}$ of
atomic propositions, and define the set of Boolean constraints $\BC(\AP)$
as the set of all propositional formulas built on~$\AP$.

\begin{definition}
  A \newdef{timed automaton} is a
  tuple $\calT=\tuple{S, S_0, AP, \Cl, T,  F}$ where
  $S$ is a finite set of states, 
  $S_0 \subseteq S$ is a set of initial states,
  $AP$~is a finite set of atomic propositions,
  $\Cl$~is a finite set of clocks,
  $T \subseteq S \times \BC(AP) \times \ClC(\Cl) \times 2^\Cl \times
  S$ is a finite set of transitions,
  and $F \subseteq S$ is the set of accepting states.
\end{definition}

   We distinguish the following classes of timed automata.
   A \emph{safety} timed automaton is such that there are no transitions
   from~$S \setminus F$ to~$F$.
   Conversely a~\emph{co-safety} timed automaton is such that there are no transitions from~$F$ to~$S\setminus F$.

For a transition~$t=(s,c,g,r,s')\in T$ of a timed automaton, we~define
${\src(t)=s}$, ${\tgt(t)=s'}$, ${\bool(t)=c}$, ${\guard(t)=g}$, and
${\reset(t)=r}$.
Note that guards are pairs of Boolean and timed guards that can be
interpreted (and will be noted) as conjunctions since the two types of
guards do not interfere.

With~a timed automaton~$\calT$, we~associate the
infinite-state automaton
$\calS(\calT)=\tuple{Q,Q_0, \Sigma, D, Q_F}$ that defines its semantics, where
\begin{itemize}
\item the set of states $Q$ contains all {\em configurations}~$(s,v) \in S
  \times \bbN^\Cl$;
\item the initial states are obtained by adjoining the null valuation
  (all~clocks are mapped to~zero) to initial states $S_0$, i.e. $Q_0=
  S_0 \times \bfzero$;
  \item $\Sigma= 2^{AP}$ is the alphabet of actions, i.e. valuations
    of all Boolean variables;
  \item transitions in~$D$ are combinations of a transition of the
    TA and a one-time-unit delay. Formally, given a
    letter~$\sigma\in\Sigma$ and two configurations $(s,v)$ and
    $(s',v')$, there is a transition $((s,v),\sigma,(s',v'))$ in~$D$
    if, and only~if, there is a transition~$(s,c,g,r,s')$ in~$T$ such
    that
    $\sigma\models c$ and $v\models g$, and 
    ${v'=(v[r\leftarrow 0])+1}$.
 \item  $Q_F = F \times \bbN^\Cl$ is the set of accepting configurations.
\end{itemize}

Our semantics thus makes it compulsory to alternate between taking a
transition of the~TA (possibly a self-loop) and taking a one-time-unit
delay. Self-loops can be used to emulate invariants in states.

The transition system~$\calS(\calT)$ is infinite because we impose no
bound on the values of the clocks during executions. However, as
in the setting of TA~\cite{AD90}, the
exact value of a clock is irrelevant as soon as it exceeds the largest
integer constant with which it is compared. We~could thus easily
modify the definition of~$\calS(\calT)$ in such a way that it only
contains finitely many states.

A~\emph{run} of~$\calT$ is a run of its associated infinite-state
automaton~$\calS(\calT)$. It~can be represented as a
sequence along which configurations and actions alternate:
\(  {(s_0,v_0)\cdot \sigma_1 \cdot
    (s_1,v1) \cdot \sigma_2 \cdots (s_n,v_n)\cdots}.
  \)
  A finite run is accepted  if it ends in $Q_F$.

  A~\emph{trace} of a run is its projection on the set of actions.
  In~other terms, it is a finite or infinite sequence $\sigma = (\sigma_i)_{0\leq
  i< l}$ of actions
where $l\in\bbN\cup\{+\infty\}$ is the
length of~$\sigma$, denoted by~$\size\sigma$.
Finite traces
belong to $\Sigma^*$ and infinite ones to $\Sigma^\omega$.
A~finite trace is accepted by~$\calT$
if a run on that trace is accepted.
We note $\Tr(\calT)$ the set of accepted traces.
For $P\subseteq Q$  we will also note $\Tr_P(\calT)$ the set of traces of runs ending in $P$.

  Consider the following sets, where~$F$ is an atomic proposition denoting~$Q_F$:
  \begin{itemize}
  \item $\success_\calT= F \et \A\G F$: 
    accepting configurations from which non-accepting configurations are unreachable are called \emph{success};
    notice that it is  impossible to escape from $\success_\calT$ since $\success_\calT \implies  \A\G \;\success_\calT$;
  \item $\error_\calT=\neg F \et \A\G \neg F$: 
    non-accepting configurations from which accepting configurations are unreachable are called {\em error};
    notice also that  it is impossible to escape from $\error_\calT$ since $\error_\calT \implies \A\G \;\error_\calT$; 
    \end{itemize}

  Note that 
  in safety TAs, $\neg F \implies \A \G \neg F$ since it is impossible to escape from the set of non-accepting configurations, thus   $\error_\calT=\neg F$; 
  symmetrically in co-safety~TAs, $F \implies \A\G\;F$
  since it is impossible to escape from the set of  accepting configurations, thus $\success_\calT= F$.

We require that our TAs are {\em complete},
meaning that from any (reachable) configuration~$(s,v)$, and for any
subset~$b$ of~$\AP$, there is~$t=(s,c,g,r,s') \in T$ such that $b\models c$ and $v\models g$. 
This is no loss
of generality since
missing transitions can be directed to a trap state,
and self-loops can be added to allow time
elapse.

The TAs that we consider are also \emph{deterministic}:
for any two transitions $(s,c_1,g_1,r_1,s_1)$
and $(s,c_2,g_2,r_2,s_2)$ issued from a same source~$s$,
if both $c_1\et c_2$ and $g_1\et g_2$ are satisfiable, then $s_1=s_2$ and
$r_1=r_2$.
Examples of complete, deterministic TAs expressing requirements
are depicted on Fig.~\ref{fig-exTA-SUP}, in Example~\ref{ex-rtconsist}.

We consider the product of timed automata, as follows:
\begin{definition}
  Given two TAs
$\calT_1=\tuple{S_1, S_{1,0}, AP_1, \Cl_1, T_1, F_1}$ and
$\calT_2=\tuple{S_2, S_{2,0}, AP_2, \Cl_2, T_2, F_2}$ with disjoint
clock sets (i.e., $\Cl_1 \cap \Cl_2=\emptyset$), their {\em product}
$\calT_1 \otimes \calT_2$ is a TA $\calT=\tuple{S, S_0,
  AP, \Cl, T, F}$ where $S=S_1 \times S_2$, $S_0=S_{1,0} \times S_{2,0}$,
$AP=AP_1 \cup AP_2$, $\Cl=\Cl_1 \cup \Cl_2$,
$F= F_1 \times F_2$
and the set of transitions is defined as follows:
there is a transition $((s_1,s_2), c,g,r, (s'_1,s'_2))$ in~$T$ if there
are transitions $(s_1, c_1,g_1,r_1, s'_1)$ in $T_1$ and
$(s_2, c_2, g_2, r_2, s'_2)$
in $T_2$ with $c=c_1 \et c_2$,
$g=g_1 \et g_2$,
and
$r=r_1 \cup r_2$.
\end{definition}

Note that completeness and determinism are preserved by product.
The~product of~TAs can be generalized to
an arbitrary number of TAs:
for a set $\calR= \{R_i\}_{i\in I}$ of requirements, each specified by
a TA $\calT_i(R_i)$, we note $\otimes \calR$ the
requirement specified by the TA $\otimes_{i\in I}
\calT_i(R_i)$.

Note that in this definition, clocks of factor automata are disjoint, 
while atomic propositions are not, which may cause conflicts in guards of the product, and possibly inconsistencies as will be seen later. 
Also note that the product of two automata  visits its accepting states if both automata do ($F= F_1\et F_2$),
while by complementation it visits non-accepting states if one of the automata does
($\neg F 
= \neg F_1 \vee \neg F_2$).
For the product automaton, we directly define (without relying on $F$) 
$\success_\calT=\success_{\calT_1} \et \success_{\calT_2}$
and $\error_\calT= \error_{\calT_1} \vee \error_{\calT_2}$, and both are trap sets.
The definitions of $\error$ and $\success$ thus depend on the context:
these are defined by the formulas
$\neg F_i \et \A\G \neg F_i$ and $F_i \et \A\G F_i$ for
the TAs $\calT_i$ representing the given requirements;
for the \emph{products} of these automata, $\error_\calT$ (resp. $\success_\calT$)  is the disjunction (resp. conjunction) of $\error_{\calT_i}$ (resp. $\success_{\calT_i}$) of their operands.
Notice that we have $\success_\calT = F \et \A\G F$, but only $\error_\calT \subseteq \neg F\et \A\G \neg F$.
The inclusion is in general strict, but becomes an equality when both $\calT_1$ and $\calT_2$ are safety TAs.

For the rest of this document, we consider complete deterministic
timed automata  (CDTAs for short) with accepting states $F$.

\subsection{Timed automata as requirements}
\label{sec:input}

We use complete deterministic TAs to encode requirements and identify the requirements with the CDTAs that define them.
Remember that $\error$ (resp. $\success$) are sets of configurations from which one cannot escape. 
Intuitively, entering an $\error$ (resp. $\success$) configuration of a CDTA  
corresponds to violating (resp. satisfying) the corresponding requirement definitively:

  \begin{definition}
    For any requirement $R$ defined by a complete deterministic timed automaton~and any finite or infinite
    trace~$\sigma$, we write
    $\sigma \;\fails \; R$ if running  $\sigma$ in $R$ enters  $\error_R$,
    and write $\sigma \;\succ \; R$ if it  enters $\success_R$.
   \end{definition}

      Note that for a finite trace $\sigma$,
      it could be the case that it does not hit $\error_R$ (resp. $\success_R$)
      but all infinite continuations inevitably~do.
      We are particularly interested in such cases; we thus
      define the following notations
  for finite traces:

  \begin{definition}
    For a finite trace $\sigma$, and a requirement  $R$ defined by a  CDTA,
     we write $\sigma \;\Ifails \;R$ if for all infinite traces $\sigma'$, 
     $\sigma\cdot \sigma' \;\fails \; R$.
     Similarly $\sigma \;\Isucc \; R$ if for all infinite traces $\sigma'$,
     $\sigma\cdot \sigma' \;\succ \; R$.
   \end{definition}

  Clearly, for finite traces, 
  $\fails$ (resp. $\succ$) is stronger than $\Ifails$ (resp. $\Isucc$).
  Indeed   $\sigma \;\fails \; R$  ($\sigma \;\succ \; R)$
    means reaching a configuration in $\error_R$ (resp. $\success_R$),   
  while  $\sigma \;\Ifails \; R$  ($\sigma \;\Isucc \; R)$
  means reaching a configuration in $\A\F \;\error_R$ (resp. $\A\F \;\success_R$).   
  And $\error_R$ implies $\A\G \;\error_R$, which implies $\A\F \;\error_R$
  (and similarly for $\success_R$).

  For a given trace~$\sigma$, and \emph{set} of timed automata~$\calR=\{\calT_i\}_{i \in I}$,
  we write~$\sigma \;\fails \;\calR$ (resp $\sigma \;\succ \;\calR$)  
  to mean that~$\sigma \;\fails \;\otimes \calR$  (resp. $\sigma \;\succ \;\otimes \calR$).
  Note the following simple facts:
  given $\calR' \subseteq \calR$, 
  for any finite trace $\sigma$,
  if $\sigma \;\fails \;\calR'$ then $\sigma \;\fails \;\calR$,
  and if $\sigma \;\Ifails \;\calR'$ then $\sigma \;\Ifails \;\calR$,
  while conversely,
  if $\sigma \;\succ \;\calR$ then  $\sigma \;\succ \;\calR'$,
   and if $\sigma \;\Isucc \;\calR$ then $\sigma \;\Isucc \;\calR'$.

\paragraph{Simplified Universal Patterns (SUP).}
  TAs can be used to express the semantics of Simplified Universal
  Pattern (SUP)~\cite{TeigeBH16,Becker2019}, a pattern language that
  is used to define requirements.  Compared to TAs, SUPs offer a more
  intuitive but less expressive way of writing requirements.  Since
  partial consistency was introduced for SUP, we briefly introduce
  them. 
  An SUP has the following form:
  \[
(\TSE, \TC, \TEE)[\Tmin, \Tmax]  
  \xrightarrow{[\Lmin, \Lmax]}
(\ASE, \AC, \AEE)[\Amin, \Amax],
\]
where \TSE, \TC, \TEE,  
\ASE, \AC, \AEE,  
are Boolean formulas on a set~$\AP$ of atomic propositions,
\Tmin, \Tmax,  \Lmin, \Lmax, \Amin, \Amax
are integers.

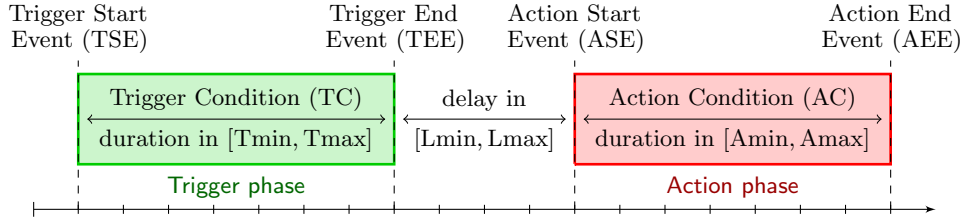
\begin{figure}[ht]
  \centering
  \begin{tikzpicture}[scale=.6]
    \fontsize{9pt}{10pt}\selectfont
    \draw[|-latex'] (-1,-1) -- +(20,0);
    \foreach \i in {0,...,18}{\draw (\i,-.9) -- +(0,-.2);}
    \draw[vert,line width=1pt] (0,0) -| (7,2) -| cycle;
    \path (3.5,-.5) node {\sffamily\textcolor{green!40!black}{Trigger phase}};
    \path (7,3) node {\begin{minipage}{2cm}\centering
        Trigger End\\ Event (TEE)\end{minipage}};
    \path[use as bounding box] (0,0);
    \path (0,3) node {\begin{minipage}{2cm}\centering
        Trigger Start\\ Event (TSE)\end{minipage}};
    \draw[<->] (0.2,1) -- (6.8,1)
    node[midway,above] {Trigger Condition (TC)}
    node[midway,below] {duration in $[\Tmin,\Tmax]$};
    \draw[dashed] (0,-1) -- (0,2.3);
    \draw[dashed] (7,-1) -- (7,2.3);
    \draw[rouge,line width=1pt] (11,0) -| (18,2) -| cycle;
    \path (14.5,-.5) node {\sffamily\textcolor{red!60!black}{Action phase}};
    \path (11,3) node {\begin{minipage}{2cm}\centering
        Action Start\\ Event (ASE)\end{minipage}};
    \path (18,3) node {\begin{minipage}{2cm}\centering
        Action End\\ Event (AEE)\end{minipage}};
    \draw[<->] (11.2,1) -- (17.8,1)
    node[midway,above] {Action Condition (AC)}
    node[midway,below] {duration in $[\Amin,\Amax]$};
    \draw[dashed] (11,-1) -- (11,2.3);
    \draw[dashed] (18,-1) -- (18,2.3);
    \draw[<->] (7.2,1) -- (10.8,1)
    node[midway,above] {delay in}
    node[midway,below] {$[\Lmin,\Lmax]$};
\end{tikzpicture}
  \caption{Intuitive semantics of SUP}\label{fig-SUP}
\end{figure}
Figure~\ref{fig-SUP} illustrates the intuitive semantics of SUP.
A~{\em trigger phase} (left) is realized,
if \TSE
is confirmed within a duration in $[\Tmin;\Tmax]$,
that is, if \TC holds until \TEE occurs;
otherwise the trigger is \emph{aborted}.
For the SUP instance to be satisfied,
following each realized trigger phase, 
an \emph{action phase} must be realized: an~action phase starts with \ASE
within $[\Lmin;\Lmax]$ time units after the end of the trigger phase,
and then \AC must hold until \AEE occurs within $[\Amin,\Amax]$ time units.
Otherwise, the~SUP is \emph{violated}.
Following~\cite{Becker2019}, one can translate SUP instances (and repetitions of them) into complete deterministic timed automata.
In fact all SUPs can be written as safety or co-safety CDTAs.

\begin{example}\label{ex-rtconsist}
    Consider the following two SUPs:
    $R_1: \request \xrightarrow{[3;4]} \response$,
    and $R_2: \repair  \xrightarrow{[5;5]} \neg\response[3;3]$,
    where an SUP of the form 
    $(p,p,p)[0;0] \xrightarrow{[0;1]} (q,q,q)[0;0]$
    is written $p \xrightarrow{[0;1]} q$.

    The first requirement models a system that has to respond to any
    request within 3 to 4 time units. The second requirement states that
    if the system enters a maintenance phase, then it will be off (and
    cannot respond) after 5 time units, and for a duration of 3
    time units.
    Figure~\ref{fig-exTA-SUP} displays the (safety) automata encoding these two SUPs 
    where $E_i$ states are non-accepting trap states and all other ones are accepting.

    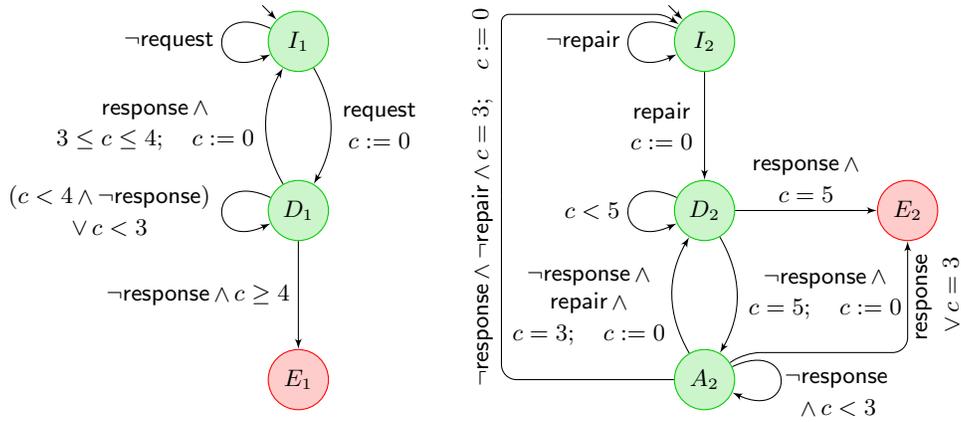
\begin{figure}[ht]
          \centering
          \begin{tikzpicture}[scale=0.9]
        \path[use as bounding box] (-1,-5.3) -- (7,.3);
        \begin{scope}
          \draw (0,0) node[rond7,minimum size=8mm,vert] (i1) {$I_1$};
          \draw[latex'-] (i1.135) -- +(135:3mm);
        \draw (0,-2.5) node[rond7,minimum size=8mm,vert] (d1) {$D_1$};
        \draw (0,-5) node[rond7,minimum size=8mm,rouge] (e1) {$E_1$};
        \draw (i1) edge[out=155,in=-155,looseness=7,-latex']
        node[left] {$\neg\textsf{request}$} (i1);
        \draw (d1) edge[out=155,in=-155,looseness=7,-latex']
        node[left] {$\gf{(c<4 \et \neg\textsf{response})}{\ou c<3}$}
        (d1);
        \draw (i1) edge[-latex',bend left] node[right] {$\gf{\textsf{request}}
          {c:=0}$} (d1);
        \draw (d1) edge[-latex',bend left] node[left] {$\gf{\textsf{response}
            \et}{ 3\leq c\leq 4;\quad c:=0}$} (i1);
        \draw (d1) edge[-latex'] node[left]
          {$\neg\textsf{response} \et c\geq 4$} (e1);
        \end{scope}
        \begin{scope}[xshift=6cm]
          \draw (0,0) node[rond7,minimum size=8mm,vert] (i2) {$I_2$};
          \draw[latex'-] (i2.135) -- +(135:3mm);
          \draw (0,-2.5) node[rond7,minimum size=8mm,vert] (d2) {$D_2$};
          \draw (0,-5) node[rond7,minimum size=8mm,vert] (a2) {$A_2$};
          \draw (3,-2.5) node[rond7,minimum size=8mm,rouge] (e2) {$E_2$};
          \draw (i2) edge[out=155,in=-155,looseness=7,-latex']
          node[left] {$\neg\textsf{repair}$} (i2);
          \draw (d2) edge[out=155,in=-155,looseness=7,-latex']
          node[left] {$c<5$}   (d2);
          \draw (i2) edge[-latex'] node[left] {$\gf{\textsf{repair}}
            {c:=0}$} (d2);
          \draw (d2) edge[-latex',bend left] node[right]
                {$\gf{\neg\textsf{response}
              \et}{c=5;\quad c:=0}$} (a2);
          \draw (a2) edge[out=25,in=-25,looseness=7,-latex']
          node[right,pos=.7] {$\gf{\neg\textsf{response}}
            {{}\et c<3}$} (a2);
          \draw[rounded corners=2mm] (a2) -| +(-3,2.7) node[coordinate] (int) {}
          node[pos=1,above,sloped] {$\neg\textsf{response}\et \neg\textsf{repair}
            \et c=3;\quad c:=0$};
          \draw[rounded corners=1mm,-latex'] (int) |- ($(i2)+(-.6,.4)$) -- (i2);
          \draw (d2) edge[-latex'] node[above] {$\gf{\textsf{response} \et}
          {c=5}$} (e2);
          \draw[rounded corners=2mm,-latex'] (a2) -- +(.6,.4) -| (e2)
            node[pos=.75,below,sloped] {$\gf{\textsf{response}}{{} \ou c=3}$};
          \draw (a2) edge[-latex',bend left] node[left=-1mm,pos=.4]
                {$\gf{\neg\textsf{response} \et}{
                    \gf{\textsf{repair} \et}{c=3;\quad c:=0}}$} (d2);
        \end{scope}
          \end{tikzpicture}
      \caption{Timed automata encoding $R_1$ and~$R_2$}\label{fig-exTA-SUP}
    \end{figure}
    \end{example}

\subsection{Consistency notions}
\subsubsection{RT-consistency.}
\label{sec:rt-consistency}

We reformulate the original rt-consistency notion, introduced
in~\cite{Post2011}.

\begin{definition}
  \label{def:rt-inconsistency}
  Let~$\calR$ be a set of requirements. Then~$\calR$ is
  \emph{rt-consistent}
  if, and only if,
  for all finite traces~$\sigma$, 
  if $\sigma \;\Ifails \;\calR$ then $\sigma \;\fails \;\calR$.
\end{definition}

Thus the set~$\calR$ is rt-consistent 
if any finite trace that inevitably fails,
immediately fails. 
This is indeed equivalent to the formulation in~\cite{Post2011},
which says that all finite
traces not violating any requirement
can be extended to an infinite trace
not violating any of them
(i.e. ~$\neg(\sigma \;\fails \;\calR)$ implies  $\neg(\sigma \;\Ifails \;\calR)$).
Notice that rt-consistency (w.r.t. $\error_\calR$) could be generalized to rt-consistency w.r.t $\success_\calR$: if $\sigma \;\Isucc \;\calR$ then $\sigma \;\succ \;\calR$;
and all following results easily generalize to rt-consistency w.r.t. $\success_R$ with similar treatment.

Observe that even when all individual requirements are rt-consistent
(i.e., for all $R \in \calR$ and all traces $\sigma$, it~holds $\sigma \;\Ifails \; R \implies \sigma \;\fails \; R$)
their conjunction (i.e. the product $\otimes \calR$)  may not be rt-consistent;
for~instance, taken individually, both requirements~$R_1$ and~$R_2$ of
Example~\ref{ex-rtconsist} are rt-consistent, but their product is not,
as explained in Example~\ref{Ex-rt-inconsistent}).
Rt-consistency requires that $\fails$ and~$\Ifails$
be equivalent for all traces in the product automaton.

Rather than using duration calculus as in \cite{Post2011},
we show that \CTL model checking can be used in a discrete-time setting to check rt-consistency.
In~\CTL, rt-consistency of~$\calR$ can be expressed by requiring $\A \F \;\error_\calR \Leftrightarrow  \error_\calR$ at all reachable states.
Since $\error_\calR$ is absorbing,
a trace ending in a configuration
in $\neg \error_\calR \et \A\F \;\error_\calR$
is a \emph{witness to rt-inconsistency.}
Moreover, only configurations in $\neg \error_\calR$ need to be traversed to
reach such configurations;
and such a configuration exists if, and only~if, configurations exist in $\neg \error_\calR$ with all immediate successors in $\error_\calR$, i.e., $\A\X \;\error$ is true.
In fact, we obtain the following property.
\begin{restatable}{lemma}{lemmartformula}
  \label{lemma:rt-formula}
  A given set of requirements~$\calR$ has a witness to
  rt-inconsistency if, and only~if, $\calR \models \E(\neg
  \error_\calR \; \U \; (\neg \error_\calR \et \A\X
  \;\error_\calR))$.
\end{restatable}

\begin{example}\label{Ex-rt-inconsistent}
  The requirements in Example~\ref{ex-rtconsist}  are not rt-consistent: 
  consider a finite
  trace~$\sigma$ where the \repair signal is received, followed 3 time
  units later with a \request.  Then $\neg (\sigma \;\fails \; R_1 \land
  R_2)$; the joint run of the automata are as follows:
  \begin{multline*}
    \def\cc#1#2{\genfrac{}{}{0pt}{1}{c_1=#1}{c_2=#2}}
    (I_1,I_2,\cc00) \xrightarrow[\scriptstyle (+\text{delay})]{\repair}
  (I_1,D_2,\cc11) \xrightarrow[\scriptstyle (+\text{delay})]{\star}
  (I_1,D_2,\cc22) \\
    \def\cc#1#2{\genfrac{}{}{0pt}{1}{c_1=#1}{c_2=#2}}
    \xrightarrow[\scriptstyle(+\text{delay})]{\star}
  (I_1,D_2,\cc33)
  \xrightarrow[\scriptstyle(+\text{delay})]{\request}
  (D_1,D_2,\cc14).
  \end{multline*}
  From this last configuration, it~can be checked that no
  continuations of this trace will avoid reaching~$E_1$ or~$E_2$:
  indeed, both automata will first
  loop in their current states~$D_1$ and~$D_2$, reaching configuration
  $(D_1,D_2), c_1=2, c_2=5$. In~order to avoid visiting~$E_2$, the
  next two steps must satisfy~$\neg\textsf{response}$, thereby
  reaching  $(D_1, A_2), c_1=4, c_2=2$. 
  From there, we~have a conflict:
  if~\textsf{response} is true at the next step, $R_2$ reaches~$E_2$,
  while if \textsf{response} is false, $R_1$ reaches $E_1$.
  
  Now, assume we add the following requirement, which expresses that no
  request can be received during maintenance: $R_3: \repair \xrightarrow{} \neg\request[5;5]$.
  This rules out the above trace, and it can be checked that the
  resulting set of requirements is now rt-consistent.
\end{example}

\subsubsection{Partial consistency.}

\emph{Partial consistency} was introduced in~\cite{Becker2019} as
an alternative, more efficient check to detect inconsistencies in SUP
requirements.
We~here generalize this notion to CDTAs.
The~name \emph{partial consistency} might be misleading since it does not directly compare with rt-consistency:
partial inconsistency identifies risky situations for pairs of requirements that could cause rt-inconsistency of the whole set.
In~this paper, we~formalize this link, and show how 
to lift witnesses of partial inconsistencies to witnesses of rt-inconsistencies.

In a requirement $R_i$, let us call \emph{action} configurations those
configurations allowing to enter immediately
$\error_{R_i}$ (i.e. satisfying $\E\X \;\error_{R_i}$)\footnote{For SUPs, such configurations correspond to  \emph{action} phases, hence the name.}.
Then, action configurations
that have an infinite continuation that avoids $\error_{R_i}$ are characterized by
$\E\X \;\error_{R_i} \et \neg \A\F \;\error_{R_i}$.
Now, $\E\X \;\error_{R_1} \et \E\X \;\error_{R_2}$
means we are simultaneously at action configurations of both $R_1$ and
$R_2$.  In this case, even though there are separate continuations
that avoid $\error_{R_1}$ and~$\error_{R_2}$, there may not be a
common one. In our generalization of partial consistency,
we focus our attention to checking that a common continuation
exists for this type of configurations which are seen as ``risky''
since they are in the proximity of error.

  Let $\reach_k(\calR)$ denote the set of configurations of $\calR$ reachable within $k$~steps.%
  \begin{definition}
  \label{def:partial_b}
  Consider requirements~$R_1,R_2$ and a set~$\calR'$ of requirements.
  We~say that~$R_1$ and~$R_2$ are \emph{partially consistent w.r.t. $\calR'$} if  for all $k \in \bbN$,
  \begin{multline}
      \label{eqn:partial-consist}
      \text{if, for all~$i\in\{1,2\}$,} \\
      \shoveleft{\exists s_i \in \reach_k(\calR_1\times \calR_2 \times \calR').\ s_i \models \E\X \;\error_{R_1} \et \E\X \;\error_{R_2} \et}\\
      \shoveright{ \neg \A\F (\error_{\calR'} \vee \error_{R_i})}
      \\[-3mm]
      \shoveleft{\text{then}}\\
      \shoveleft{\exists s \in \reach_k(\calR_1\times \calR_2 \times \calR').\ s \models \E\X \;\error_{R_1} \et \E\X \;\error_{R_2} \et }\\
      \neg \A\F(\error_{\calR'}\vee\error_{R_1}\vee \error_{R_2}).
        \end{multline}
  \end{definition}

  Partial consistency requires that for all depths $k$, if infinite traces for both requirements can be found leading to an action configuration within $k$ steps, and neither violate the requirement itself nor $\calR'$, then  a single infinite trace must exist that reaches action configurations of both requirements within $k$ steps, and does not violate any of them, nor $\calR'$.
  Therefore, a witness of partial inconsistency is a number $k\geq 0$ and two infinite sequences $\sigma_1$ and $\sigma_2$
  such that, $\sigma_i$ reaches actions phases of both requirements within $k$ steps, and never fails~$R_i$ or~$\calR'$,
  such that 
  there are no infinite traces that do so without violating one of the requirements $R_1$, $R_2$ or~$\calR'$.

We establish that partial consistency is a necessary condition for the rt-consistency of
the \emph{subset} $\calR'\cup\{R_1,R_2\}$, since
counterexamples for the former provide counterexamples for the latter:
\begin{restatable}{lemma}{lemmapartrt}
  \label{lemma:partial-inconsistent-rt-inconsistent}
  If~$R_1$ and~$R_2$ are partially inconsistent w.r.t. $\calR'$, 
  then $\calR' \cup\{R_1,R_2\}$ is rt-inconsistent.
\end{restatable}

To efficiently find counterexamples to partial consistency,
we consider the following approximation,
which is similar to that of \cite{Becker2019}
but generalized to CDTAs.
Given bounds~$\alpha,\beta > 0$, requirements~$R_1,R_2$  are \emph{$(\alpha,\beta)$-bounded partially consistent} if  for all $k \leq \alpha$,
  \begin{multline}
      \label{eqn:partial-consist-approx}
      \text{if, for all~$i\in\{1,2\}$,} \\
      \shoveleft{\exists s_i \in \reach_k(\calR_1\times \calR_2 \times \calR').\ s_i \models \E\X \;\error_{R_1} \et \E\X \;\error_{R_2} \et}\\
      \shoveright{ \neg \A\F_{\alpha-k} (\error_{\calR'} \vee \error_{R_i})}
      \\[-3mm]
      \shoveleft{\text{then}}\\
      \shoveleft{\exists s \in \reach_k(\calR_1\times \calR_2 \times \calR').\ s \models \E\X \;\error_{R_1} \et \E\X \;\error_{R_2} \et }\\
      \neg \A\F_{\alpha+\beta-k}(\error_{\calR'}\vee\error_{R_1}\vee \error_{R_2}).
        \end{multline}
where $\A\F_l \phi$ means the inevitability of $\phi$ within $l$ steps,
which can be expressed in \CTL as the disjunction of all formulas of the form $\A\X(\phi \vee \A\X(\cdots \phi\vee \A\X \phi))$ with
$l$ repetitions of~$\A\X$.
Thus the approximation consists in looking for witnesses of bounded length for the satisfaction of the Equation~\ref{eqn:partial-consist}).
But notice that witnesses of failure of Equation~\ref{eqn:partial-consist-approx}
are not witnesses of failure of Equation~\ref{eqn:partial-consist} which require infinite traces (see~below).

\begin{example}
  We consider again the requirements of Example~\ref{ex-rtconsist}.
Requirements~$R_1$ and~$R_2$ are not partially
consistent under empty~$\calR'$:
as soon as a trace reaches  action configurations of both requirements,
error states of any of them can be avoided, but not both of them.
Under requirement~$R_3$,
requirements $R_1$ and~$R_2$ cannot reach their action phases simultaneously, so
that with~$\calR'=\{R_3\}$, those two requirements are
partially consistent.
\end{example}

There are a few differences with the original definition of partial consistency of~\cite{Becker2019}.
First, partial consistency of~\cite{Becker2019}
only checks the very first trigger of the traces.
Moreover, it focuses on situations where,
after respective triggers, no timing allows requirements to avoid being simultaneously in action phases.
In our case,
$\E\X \;\error_{R_1} \et \E\X \;\error_{R_2}$ does not restrict simultaneous action phases to such particular ones. 
Thus we can detect more subtle inconsistencies.

The second difference is that the bounded approximation in \cite{Becker2019} checks for the existence
of a lasso-shaped execution in the automata that recognize the SUP requirements.
The advantage of this is that such a lasso describes an infinite execution,
so if partial consistency holds,
so does the bounded approximation; 
while the converse is not true.
In other terms, a  witness for bounded partial inconsistency is a witness for partial inconsistency. 
In our case, we do not look for a lasso in the premise of~(\ref{eqn:partial-consist-approx}),
so this implication does not hold. We do prove, on the other hand, that rt-consistency
implies~(\ref{eqn:partial-consist-approx}); see Lemma~\ref{lemma:partial-to-rt}.

Third, in~\cite{Becker2019}, $\calR'$ contains only a specific type of requirements called invariants.
In~our case, $\calR'$~is an arbitrary subset of the requirement set.

\section{Incremental algorithms}
\label{sec:algorithms}
We provide three incremental methods to check rt-consistency of a given set
of requirements~$\calR$.  The first one provides strong guarantees
and can assess the rt-consistency of the whole set~$\calR$, or that of
its subsets, and uses \CTL model checking.
The~second one uses SAT\slash SMT solving and scales to larger~sets.
It~can \emph{detect} rt-inconsistencies
of~$\calR$, but cannot \emph{prove} rt-consistency; it~can only ensure partial
consistency.
The third one can quickly find rt-inconsistencies.

In all algorithms we consider a set~$\calR=\{R_i\}_{i \in I}$ of requirements, each given as a CDTA, and their product $\otimes\calR$.

\subsection{Incremental rt-consistency checking}

In this section, we present our incremental algorithm for
rt-consistency checking.  Unlike the previous work of~\cite{Hoenicke2006},
which uses duration calculus~\cite{Zhou1991}, our algorithm is based on
\emph{computation tree logic} (\CTL) model checking.
Rt-inconsistency of~$\calR$ reduces to checking whether a finite trace exists along which $\error_\calR$ remains false such that,
from the last configuration, $\error_\calR$ is inevitable.
Such a finite trace~$\sigma$ is called a \emph{witness} for the rt-inconsistency of~$\calR$.
Remember that, by Lemma~\ref{lemma:rt-formula}, this can be written in \CTL as
$\E[ \lnot\error_\calR \;\U\; (\lnot\error_\calR \land \A\X\; \error_\calR) ]$
 to be checked in~$\otimes\calR$.

 When the size of~$\calR$ is too large for model-checking tools to
handle, we~consider subsets $\calR'$ of~$\calR$.  
Such incomplete checks alone do not provide any guarantee;
indeed if $\calR' \subseteq \calR$,
consistency of~$\calR$ does not imply consistency of~$\calR'$, nor the opposite.
Nevertheless, they can be used to detect
rt-inconsistencies with an additional check:
\begin{restatable}{lemma}{lemmartwitness}
  \label{lemma:rt-witness-generalizes}
  Let $\sigma \in \Sigma^*$ be a witness for the rt-inconsistency of~$\calR' \subseteq \calR$.
  If~${\neg (\sigma\;\fails \;\calR)}$,
  then $\sigma$ is also a witness for the rt-inconsistency
  of~$\calR$.
\end{restatable}

\begin{algorithm}[t]
  \DontPrintSemicolon
  \KwIn{A set~$\calR$ of requirements given as CDTAs, $2 \leq n \leq |\calR|$}
  $\phi(\calR) \gets \E[ \lnot\error_\calR ~\U (\lnot\error_\calR \land \A\X~\error_\calR) ]$\;
  \For{\textrm{\bf all} pairs $\{R_1,R_2\} \subseteq \calR$} {
    $\calR' \gets \{R_1,R_2\}$\;
    \While{$|\calR'| \leq n$ \textbf{ and } $\calR' \models \phi(\calR')$}{
        $\sigma \gets$ witness of~$\phi(\calR')$     \tcp{$\sigma$ witnesses rt-inconsistency of $\calR'$}    
        \eIf{$\exists R \in \calR \setminus \calR' \text{ s.t. } \sigma \; \fails \; R$}{
          $\calR' \gets \calR' \cup\{R\}$\;
        }{
          \Return $\sigma$ \tcp{$\sigma$ witnesses rt-inconsistency of $\calR$}
        }
    }
  }
  \Return $\emptyset$ \tcp{no witness for the rt-inconsistency of $\calR$ is found}\;
  \caption{Incremental rt-consistency checking algorithm.
  In order to avoid checking the same subsets of~$\calR'$ several times, one can store the subsets seen so far and break the while loop when~$\calR'$ has already been treated.
  }
  \label{algo:inc-rt-cons}
\end{algorithm}

Let us now  describe our procedure summarized in Algorithm~\ref{algo:inc-rt-cons}.
Given~$\calR$ and a bound~${n \leq |\calR|}$,
we consider subsets of~$\calR$ of size up to~$n$, starting with subsets of size~$2$.
Assume a subset~$\calR'\subseteq \calR$ is found to be rt-inconsistent with a witness trace~$\sigma$. 
We check whether~$\sigma \; \fails \; \calR\setminus \calR'$.
If this is the case,
   we select $R \in \calR \setminus \calR'$ such that $\sigma \;\fails \; R$,
  and restart the analysis with~$\calR' \cup \{R\}$. 
  Notice that if~$\calR' \cup \{R\}$ is inconsistent, then~$\sigma$
cannot be a witness trace since it violates~$R$. This ensures that a new requirement
will be added to the set at each iteration.
Otherwise, by Lemma~\ref{lemma:rt-witness-generalizes}, we conclude that $\calR$ is rt-inconsistent and $\sigma$ is a witness.
If no confirmed witnesses are found, then we stop and report that no rt-inconsistency is found.
If~$n\geq |\calR|$, then one can conclude that $\calR$ is rt-consistent; otherwise the check is incomplete.

To increase the precision (to have a better chance to detect rt-inconsistencies), one can increase the bound~$n$.
In order to reduce the number of cases to check, thus giving up on completeness, one might restrict only to some subsets, for instance making sure that each requirement is covered by at least one subset.

\subsection{Incremental partial consistency checking}

\begin{algorithm}[t]
  \DontPrintSemicolon
  \KwIn{A set~$\calR$ of requirements  given as CDTAs, parameters~$\alpha,\beta >0$}
  \For{\textrm{\bf all} pairs $\{R_1,R_2\} \subseteq \calR$} {
    $\calR' \gets\emptyset$\;
    \While{Equation \eqref{eqn:partial-consist-approx} fails}{
        $(\sigma_1,\sigma_2) \gets$ witness traces for the premise of~\eqref{eqn:partial-consist-approx} for some~$k\leq \alpha$\;
        \eIf{$\exists i\in\{1,2\}, \neg(\sigma_i \; \fails \; \calR)$}{
          \Return $\sigma_i$ \tcp{witness of rt-inconsistency of $\calR$}
        }{
          \eIf{$\calR = \calR' \cup \{R_1,R_2\} $}{
            \textbf{break} \tcp{No witness is found for this pair}
          }{
            Choose~$R \in \calR$ such that $\sigma_i \; \fails \; R$ for some~$i \in\{1,2\}$\;
            $\calR' \gets \calR' \cup \{R\}$\;
          }
        }
    }
  }
  \Return $\emptyset$ \tcp{no counterexample is found}\;
  \caption{Incremental partial consistency checking algorithm.}
  \label{algo:inc-partial-cons}
\end{algorithm}

We now present an incremental algorithm for checking partial consistency
via the \emph{bounded} partial consistency checking 
in the same vein as the previous section.

Ideally, we would like to check Equation\eqref{eqn:partial-consist-approx} for all pairs~$\{R_1,R_2\}$ of requirements with respect to~$\calR' = \calR\setminus\{R_1,R_2\}$;
in fact, considering the whole set~$\calR'$ makes sure that counterexample traces
do not trivially violate requirements.
This~is costly in general, so we will start with an empty $\calR'$ and let it grow incrementally by adding
requirements as needed.
The following lemma exhibits when such counterexamples can be lifted to witnesses of rt-inconsistency:
\begin{restatable}{lemma}{lemmajhg} 
  \label{lemma:jhg}
  Let $\sigma_1$, $\sigma_2$ and $k$ be witnesses of bounded partial inconsistency for
  $R_1,R_2 \in \calR$ and $\calR' \subseteq \calR$,
  i.e. counterexamples of Equation~\ref{eqn:partial-consist-approx}.
  If, for some $i$, $\neg(\sigma_i \;\fails \;\calR)$, then $\sigma_i$ is also a witness for the rt-inconsistency of $\calR$.
\end{restatable}

The procedure is summarized in Algorithm~\ref{algo:inc-partial-cons}. 
Given pair~$(R_1,R_2)$ and set $\calR'\subseteq \calR\setminus\{R_1,R_2\}$,
integer parameters $\alpha, \beta > 0$,
checking the $(\alpha,\beta)$-bounded partial-consistency
 consists in verifying Equation~\eqref{eqn:partial-consist-approx}.
A negative check is witnessed by some $k\leq \alpha$ and a pair of traces~$\sigma_1,\sigma_2$.
If~$\neg (\sigma_i \;\fails \; \calR)$ holds for some $i \in \{1,2\}$,
the trace is returned as a counterexample by Lemma~\ref{lemma:jhg}.
Otherwise, a~requirement~$R \in \calR$ such that~$\sigma_i \; \fails \; R$ is added to the set~$\calR'$ and the procedure is repeated.
Thus, subsequent iterations will discard~$\sigma_i$ and look for other traces.
The following lemma shows that
all counterexamples returned by Algorithm~\ref{algo:inc-partial-cons} are witnesses to rt-inconsistency: 
\begin{restatable}{lemma}{lemmapartialtort}
  \label{lemma:partial-to-rt}
  Let~$\calR$ be a set of requirements, and $\sigma$ be a finite trace returned by Algorithm~\ref{algo:inc-partial-cons}.
  Then $\sigma$ is a witness for rt-inconsistency for~$\calR$.
\end{restatable}

\subsection{Incremental partial rt-consistency checking}
\begin{algorithm}[t]
  \DontPrintSemicolon
  \KwIn{A set~$\calR$ of requirements, parameters~$\alpha >0$, $n \in [1, |\calR|]$}
  \For{\textrm{\bf all} subsets $\calS \subseteq \calR$ such that $|\calS| \leq n$} {
    $\calR' \gets \emptyset$\;
        \While{$\calS\times\calR' \models \phi_{p,\alpha}$}{
          $ \sigma \gets$ witness trace for $\phi_{p,\alpha}$
        \;
        \eIf{$\neg (\sigma \;\fails \;\calR)$}{
          \Return $\sigma$ \tcp{Counterexample for $\calR$}
        }{
          \eIf{$\calR = \calR' \cup \calS $}{
            \textbf{break} \tcp{No counterexample is found for this subset}
          }{
            Choose~$R \in \calR$ such that $\sigma \;\fails \; R$\;
            $\calR' \gets \calR' \cup \{R\}$\;
          }
        }
    }
  }
  \Return $\emptyset$ \tcp{no counterexample is found}\;
  \caption{Incremental partial rt-consistency checking algorithm.}
  \label{algo:partial-rt}
\end{algorithm}

We now propose an algorithm for rt-consistency checking, that combines
an incremental approach targeting subsets of requirements (hence the
name partial), and a bounded search, providing an alternative to
Algorithm~\ref{algo:inc-rt-cons} amenable to using SMT solvers.
Intuitively, we check for the existence of configurations
where all requirements in a subset $\calS$ of $\calR$ 
\emph{immediately conflict} \textit{i.e.} $\A\X\;\error_{\calS}$,
meaning that at the next step they inevitably violate at least one requirement of $\calS$.

Let  $\calS$ be a subset of requirements of $\calR$.
We~say that $\calS$ is {\em partially rt-consistent} with respect to $\calR'$ if for all configurations $s$,
\begin{multline}
  \label{eqn:alpha-bounded-partial}
  s\models \neg\error_{\calS \cup \calR'} \implies \neg\A\X \; \error_{\calS}. \hfill
\end{multline}
This clearly implies that $\calS$ is rt-consistent, but also that no
immediate conflict affects the subset~$\calS$ in any configuration.
A witness of partial rt-inconsistency is a trace $\sigma$ that reaches a configuration $s$ satisfying $\neg\error_{\calS \cup \calR'} \et \A\X \; \error_{\calS}$.
Since $\A\X \; \error_{\calS}$  implies $\A\X \; \error_{\calR}$ (because $\error_{\calS}$ implies $\error_\calR$),
if additionally $\neg(\sigma \; \fails \;\calR)$ it is also a witness of rt-inconsistency by Lemma~\ref{lemma:rt-witness-generalizes}.
Similarly to  Lemma~\ref{lemma:rt-formula}, the existence of a witness of partial inconsistency reduces to
checking the formula
\(\phi_p= \E (\neg\error_{\calS \cup \calR'} \; \U \; (\neg\error_{\calS \cup \calR'} \et \A\X \; \error_{\calS})).\)

Partial rt-consistency can be further restricted by bounding the size of~$\calS$ and restricting the exploration depth. 
For integers $n$ and $\alpha$, we say that $\calR$ is  {\em $\alpha$-bounded $n$-partially rt-consistent} if
   Formula~\ref{eqn:alpha-bounded-partial} holds for any subset $\calS$ of size $|\calS|\leq n$, and configurations $s \in \reach_\alpha(\calR)$. 
   Checking {\em $\alpha$-bounded $n$-partial rt-inconsistency} can be done by
   replacing $\U$ by $\U_\alpha$ in $\phi_p$
   thus checking $\phi_{p,\alpha}= \E (\neg\error_{\calS \cup \calR'} \; \U_{\alpha} \; (\neg\error_{\calS \cup \calR'} \et \A\X \; \error_{\calS}))$.

   We summarize the procedure in Algorithm~\ref{algo:partial-rt},
   where, similarly to Algorithm~\ref{algo:inc-partial-cons},
   the~set~$\calR'$ is augmented by requirements failed by tentative
   counterexamples.  We~easily get the following lemma since a witness
   of $\alpha$-bounded $n$-partial rt-inconsistency that does not fail
   $\calR$ is also a witness of rt-inconsistency.
   \begin{lemma}
     \label{partial-rt-to-rt}
     Let $\calR$ be a set of requirements, and $\sigma$ be a finite trace returned by Algorithm~\ref{algo:partial-rt}.
     Then $\sigma$ is a witness for rt-inconsistency.
   \end{lemma}

\section{Preliminary Experiments}
   We experimented the different algorithms on a factory automation use case.
   In this system, a carriage and an arm cooperate to convey material: 
   objects are pushed onto the carriage,
which brings them to a position where a pushing arm places them on a conveyor belt.
The correctness of this system relies on several timed requirements 
between different elements of the system.

\looseness=-1
Table~\ref{tab:eval} shows the inconsistencies found with our algorithms on sets of requirements
of varying sizes. The largest set we considered contained 29 requirements of which 13 are
timed and the other 16 are purely Boolean. 
We~compare the incremental partial consistency and partial rt-consistency algorithms
(implemented using the SMT solver Z3~\cite{z3}),
with the incremental rt-consistency algorithm (implementing \CTL model-checking using NuSMV~\cite{nusmv}). 
Inconsistencies were detected in the first two sets, but partial consistency failed
in detecting any in set~\#2.

   These preliminary experiments show that the incremental method can help detect inconsistencies quickly.
However, since the methods are not complete, we~encourage using several algorithms in parallel.

\begin{table}[t]
  \centering
  \begin{tabular}{|l|c|c|c|c|c|} \hline
    \ \textbf{set} & \ \textbf{size} & \ \textbf{rt-consistency} \ & \ \textbf{partial consistency} \ & \ \textbf{partial rt-consistency} \ \\
    \     \ &       &  Algorithm 1       & Algorithm 2        & Algorithm 3  \\ \hline \hline    
    \ \#1 \ & 6 + 9 &  5 inconsist. (24s) &  4 inconsist. (36s) & 5 inconsist. (39s) \\
    \ \#2 \ & 8 + 10 &  1 inconsist. (21s) & \checkmark (55s) & 1 inconsist. (101s)  \\
    \ \#3 \ & 8 + 10  & \checkmark (24s) & \checkmark (61s) & \checkmark (115s) \\
    \ \#4 \ & 10 + 16 & \checkmark (359s) & \checkmark (85s) & \checkmark (141s)  \\
    \ \#5 \ & 12 + 16  & \checkmark (1143s) & \checkmark (133s) & \checkmark (227s) \\
    \ \#6 \ & 13 + 16 & \checkmark (5311s) & \checkmark (138s) & \checkmark(232s)  \\ 
    \hline
  \end{tabular}
  \medskip
  \caption{Experiments on our case study. The size shows the number of timed requirements +
  the number of (non-timed) Boolean requirements of the instance. The parameters were chosen
  as $\alpha=40$ and~$n=2$. The sign \checkmark means that no inconsistencies were found.
  The experiments were run on a 1.9Ghz processor with a timeout of 3 hours.
  }%
  \label{tab:eval}
\end{table}

\section{Conclusion}

In this paper, we studied the notions of rt-consistency and partial consistency.
We showed how to reduce the problem to \CTL model checking  on timed automata models, and presented algorithms that can detect rt-inconsistencies. Our preliminary experiments show encouraging results.
As future work, we will extensively evaluate the ability of these algorithms to
capture inconsistencies, and their performances on large realistic use cases.
One might investigate other variants of the (partial) consistency notions, with the goal of detecting more inconsistencies more efficiently.
There is a trade-off to find for such partial consistency algorithms.
In~fact, they might allow one to examine more potential counterexample witnesses, which means that one might detect more inconsistencies, but one might also have to deal with more false positives.
Another interesting question is how to correct rt-inconsistencies e.g. by adding new requirements.
\enlargethispage{2mm}
\bibliographystyle{alpha}
\bibliography{biblio}

\iffinal
\end{document}
\fi

\newpage
\appendix
\section{Proofs}

\lemmartformula*

\begin{proof}
Let us consider the following formula  \[\phi(\calR) = \E(\neg
  \error_\calR \; \U \; (\neg \error_\calR \;\land \;\A\X
  \;\error_\calR))\]
By definition, $\calR$ is rt-inconsistent if there is a reachable
configuration~$s$ such that~$s \models \lnot \error_\calR \land \A\F\; \error_\calR $.
It is thus clear that if the initial state of~$\otimes \calR$ satisfies~$\phi(\calR)$,
then $\calR$ is rt-inconsistent.

Let us assume that~$\calR$ is rt-inconsistent, and consider a reachable configuration~$s$
satisfying~$\lnot \error_\calR \land \A\F\;\error_\calR$.
Let~$\rho$ denote the run that ends in~$s$. Since~$\error_\calR$ is absorbing,
all states of~$\rho$ satisfy~$\lnot \error_\calR$.
We show that there exists some configuration~$s'$ reachable from~$s$
with both~$s' \models \lnot \error_\calR$
and  $s' \models \A\X\;\error_\calR$.
To see this, we build a run from~$s$ inductively as follows.
Initially, the run is at configuration~$s$. At any moment, if
the current configuration has a successor satisyfing $\lnot\error_\calR$,
we choose one arbitrarily and extend the run.
If there are no such successors, then this provides the configuration~$s'$
as desired. Notice that this constructed run cannot be infinite, since
this would contradict that~$s \models \A\F\;\error_\calR$, so such a~$s'$ must exist.

Now the run we obtain from the initial configuration to~$s'$ is a witness for
$\phi(\calR)$.
\end{proof}

\lemmapartrt*

\begin{proof}
  Consider $k\geq 0$, and traces~$\sigma_1,\sigma_2$ which are witnesses to partial inconsistency,
  as well as configurations~$s_i \in \reach_k(\calR_1\times \calR_2\times \calR')$.
  We have~$s_1 \models \lnot \error_{\calR_1} \land \lnot \error_{\calR_2}
  \land \lnot \error_{\calR'}$.
  Rt-consistency requires that there exists an infinite continuation from~$s_1$
  satisfying $\lnot \error_{\calR_1} \land \lnot \error_{\calR_2}
  \land \lnot \error_{\calR'}$.
  However, since~\eqref{eqn:partial-consist} does not hold,
  there is no state~$s \in \reach_k(\calR_1\times \calR_2\times \calR')$
  satisfying both~$s \models \E\X\;\error_{\calR_1} \land \E\X\;\error_{\calR_2}$ and admitting
  such an infinite continuation. Therefore, $s_i$ cannot have such a continuation,
  which proves that~$\calR'\cup\{R_1,R_2\}$ is rt-inconsistent.
\end{proof}

\lemmartwitness*
\begin{proof}
  

      
  In fact,  if $\sigma$ is a witness of rt-inconsistency in $\calR'$,  by definition
  $\neg (\sigma \; \fails \;  \calR')$ but $\sigma \; \Ifails \; \calR'$.
  Since $\calR' \subseteq \calR$,  (inevitably) failing $\calR$ implies (inevitably) failing $\calR'$
  ($\sigma \; \Ifails \; \calR'$  implies  $\sigma \; \Ifails \; \calR$).
  By hypothesis, $\neg (\sigma \; \fails\; \calR')$, but it may be the case that $\sigma ; \fails\; \calR$.
  If additionnally $\neg (\sigma \; \fails\; \calR)$, then
  we can conclude that  $\sigma$ is a witness of rt-inconsistency of $\calR$.
  \end{proof}

\lemmajhg*

  \begin{proof}
    For any $i \in \{1,2\}$, if $\sigma_i$ witnesses $(\alpha,\beta)$-bounded partial inconsistency, 
    by definition $\sigma_i$ reaches a configuration $s_i$ in $reach_k(\calR_1\times \calR_2 \times \calR')$
    satisfying $\E\X \;\error_{R_1} \land \E\X \;\error_{R_2} \land \neg \A\F_{\alpha-k} (\error_{\calR'} \vee \error_{R_i})$
    but 
    no configuration $s \in reach_k(\calR_1\times \calR_2 \times \calR')$ satisfies
    $\E\X \;\error_{R_1} \;\land \;\E\X \;\error_{R_2} \land \neg \A\F_{\alpha+\beta-k}(\error_{\calR'} \; \vee \; \error_{R_1} \;\vee \;\error_{R_2})$.
    Since $s_i \models\E\X \;\error_{R_1} \; \land \; \E\X \;\error_{R_2}$, it  satisfies 
    $\A\F_{\alpha+\beta-k}(\error_{\calR'}\vee\error_{R_1}\vee \error_{R_2})$.
    If additionally $\sigma_i$ satisfies $\neg(\sigma_i \; \fails \; \calR)$,
    since $\A\F_{\alpha+\beta-k}(\error_{R'}\vee\error_{R_1}\vee \error_{R_2})$ implies $\A\F\;\error_{\calR}$,
    then $s_i$ satisfies $\neg \error_\calR \;\land\; \A\F \;\error_\calR$, thus $\sigma_i$ is a witness for rt-inconsistency.
  \end{proof}
  
\lemmapartialtort*
  \begin{proof}
    Assume that the algorithm returned a counterexample trace~$\sigma
      \in \Sigma^*$ for the outer iteration with~$R_1,R_2 \in \calR$, and
      inner iteration $\calR' \subseteq \calR$.
      The algorithm ensures that~$\neg(\sigma \; \fails \; \calR)$ (line 6).
      We then use Lemma~\ref{lemma:jhg} to conclude that $\sigma$ is a witness for rt-inconsistency for~$\calR$. 
%
    \end{proof}

\end{document}